\documentclass[copyright,creativecommons]{eptcs}

\usepackage[utf8]{inputenc}
\usepackage[T1]{fontenc}
\usepackage[english]{babel}
\usepackage{graphicx}
\usepackage{epsfig}
\usepackage{amsmath}
\usepackage{xspace}
\usepackage{clrscode}
\usepackage[charter,cal=cmcal]{mathdesign}
\usepackage{textcomp}

\usepackage{verbatim}

\usepackage[thmmarks]{ntheorem}

\newcommand{\SAT}{\ensuremath{\text{\sc Sat}}\xspace}

\newcommand{\Dmc}{\ensuremath{\mathcal{D}}\xspace}

\newcommand{\Vmc}{\ensuremath{\mathcal{V}}\xspace}

\newcommand{\eat}[1]{}

\newcommand{\comp}{\ensuremath{\textsl{Component}}\xspace}
\newcommand{\cid}{\ensuremath{\textsl{id}}\xspace}

\newcommand{\LoCo}{\ensuremath{\mathsf{LoCo}}\xspace}
\newcommand{\confprob}{\ensuremath{\mathsf{CP}}\xspace}

\newcommand{\sort}[1]{\ensuremath{\text{\sc #1}}\xspace}

\newcommand{\interpretation}{\ensuremath{\mathcal{I}}\xspace}

\newcommand{\cex}{\ensuremath{\exists^{u}_{l}}\xspace}
\newcommand{\xcex}{\ensuremath{\exists \text{!`}^{u}_{l}}\xspace}

\newtheorem{proposition}{Proposition}	

\newtheorem{definition}{Definition}
\newcounter{ale}

\newenvironment{liste}{\begin{itemize}}{\end{itemize}}
\newcommand{\aliste}{\begin{liste} \setcounter{ale}{1}}
\newcommand{\zliste}{\end{liste}}

\title{Introducing \LoCo, a Logic for Configuration Problems\thanks{Work funded by EPSRC Grant EP/G055114/1 Constraint Satisfaction for Configuration: Logical Fundamentals, Algorithms and Complexity}}

\author{Markus Aschinger, Conrad Drescher, Georg Gottlob
\institute{Department of Computer Science, University of Oxford}
\email{firstname.lastname@cs.ox.ac.uk}
}
\def\authorrunning{M. Aschinger, C. Drescher and G. Gottlob}
\def\titlerunning{\LoCo\/: A Logic for Configuration Problems}

\begin{document}

\maketitle

\begin{abstract}
In this paper we present the core of \LoCo, a logic-based high-level representation language for expressing configuration problems. 
\LoCo shall allow to model these problems in an intuitive and declarative way, the dynamic aspects of configuration notwithstanding. 
Our logic enforces that configurations contain only finitely many components and
reasoning can be reduced to the task of model construction.
\end{abstract}


\section{Configuration Problems}

Configuration systems are one of the most successful applications of AI-techniques. 
In industrial environments, they support the configuration of complex products and, compared to manual processes, help to reduce error rates and increase throughput \cite{sabin:survey}. 
The following definition by Mittal and Frayman \cite{mittal:generic_model} describes what is typically meant by a configuration problem. 

\begin{definition}[Configuration Problem]
\label{def:cp}
{\bf Given\/:} A fixed, predefined set of components, where a component is described by a set of properties, ports for connecting it to other components, constraints at each port that describe the components that can be connected at that port, and other structural constraints, some description of the desired configuration and some criteria for making optimal selections. \\
{\bf Build\/:}
One or more configurations that satisfy all the requirements, where a configuration is a set of components and a description of the connections between the components in the set, or, detect inconsistencies in the requirements. 
\label{def:conf}
\end{definition}

In typical configuration problems, the number of components needed for a solution is unknown beforehand;
for example, for some components this number depends on the choices made for other components. 
One can think of this as of creating new components on-the-fly throughout the solving process. 
Existing knowledge representation (KR) tools able to express this dynamic aspect of configuration
require that explicit bounds on all generated components be given as well as
extensive knowledge about the underlying solving algorithms. 

In this work we introduce a purely declarative logical formalism where the KR engineer only has to specify the possible numbers of connections between any two component kinds.
From this information finite bounds on the number of components needed in a configuration are inferred ---
that is, in any model of the configuration problem the number of components used is finite.
Formally this logic is a fragment of classical First Order Logic (FO), extended by existential counting quantifiers.
We plan to eventually develop translations from the logic representation into a low-level input format for various solvers, e.g. SAT or Integer Programming.

\section{Configuration Formalisms}

Over the years several different approaches for configuration have been investigated, e.g. expert systems, rule-based systems, non-monotonic reasoning, case-based reasoning, 
description logics and constraint processing. 
A recent survey is given by Junker in \cite{junker:configuration}. 

\subsection{Constraint-Based Formalisms}

Constraint satisfaction problems (CSPs) are currently the most widely used approach for the formalisation of configuration problems. 
However, the standard CSP formulation does not feature variables or sub-CSPs that are conditionally activated depending upon the values assigned to other variables.

Hence, in the area of constraint-based configuration, a number of extensions of the traditional CSP paradigm have been developed.
In Conditional CSPs \cite{mittal:dcsp} activation constraints ensure that only a relevant subset of the variables and constraints is used for generating a solution.
In Composite CSPs \cite{sabin:composite} variables can have subproblems (sub-CSPs) as values. 
In both formalisms the number of possibly activated variables and constraints has to be defined in advance. 
Accordingly, both formalisms admit translations into classic CSPs \cite{thorstensen:capturing}. 

A Generative CSP (GCSP) \cite{stumptner:generative} allows the dynamic generation of components on demand during the search process. 
The reasoning starts from certain key components and then required auxiliary components and associated connections are incrementally added. 
No explicit bounds on the number of components have to be given and the formalism allows infinite configurations to be constructed.

\subsection{Logical Frameworks}

There have also been some previous attempts to capture configuration with logic-based formalisms. 
We recall these in some detail, as they are the starting points for our configuration logic.

Classical CSPs correspond to the fragment $\exists FO_{\wedge, +}$ of $FO_{\wedge, +}$ of FO consisting of formulae built using only existential quantification and conjunction \cite{kolaitis:cqc_csp}\/:

\begin{definition}
The logical counterpart of a CSP is defined as a pair $(\phi, \mathcal{D})$, where $\mathcal{D}$ is the \textit{constraint database}, 
i.e., the extensional representation of all the constraint relations and $\phi$ is a $\exists FO_{\wedge, +}$ sentence.
Solving the CSP corresponds to deciding whether $\mathcal{D} \vDash \phi$.
\label{def:csp_logic}
\end{definition}

In the work by Gottlob et al.\ \cite{gottlob:conditional} logical implication has been added to this formalism to express the conditional inclusion of components into configurations.
This $\exists FO_{\rightarrow, \wedge, +}$ fragment of FO is one of the starting points for our own formalism.
For example, it allows us to ask whether $\Dmc \vDash (\exists x) \text{Car}(x) \land (\text{LuxuryCar}(x)\Rightarrow\text{HasSunRoof}(x))$.
A drawback of $\exists FO_{\rightarrow, \wedge, +}$ is that explicit bounds on the number of components needed has to be given (variables have a fixed finite domain)
and that all constraints must be coded in extension in the constraint database. 

There are also two prominent formalisms based on Description Logics (DLs)\/: 
The works by McGuinness et al.\ \cite{McguinnessW98} and Klein et al.\ \cite{DFKI-TM-95-01}. 
These are the other two starting points of our formalism.

In both works valid configurations are described using DL axioms. 
DLs are fragments of FO based on unary and binary predicates, so-called concepts and roles.
In both approaches concepts are used for describing components and attributes; roles are used to describe the relations between components and also between components and attributes.
Klein et al.\ reduce the task of finding a valid configuration to the problem of constructing a {\em finite} model of the axioms. 
McGuinness et al.\ propose an interactive approach where (1) the knowledge engineer adds atomic propositions to the axioms and (2) the inference engine computes the consequences until (3) eventually a finite model is obtained. 
The DLs from both formalisms always admit both finite and infinite models; hence no explicit bound on the number of components has to be given.
The absence of predicates of arity greater than two can make domain encodings unnecessarily complex.

Finally, in \cite{FriedrichStumptnerConfWS99} a logic-based formulation of GCSPs has been given; 
this formulation does not require that bounds on the component numbers be given, but admits infinite configuration models.

\section{The \LoCo Formalism}

We now introduce the core of \LoCo, a new logic-based framework for modelling practical configuration problems. 
In this work we do not yet address ports or optimal configurations.
The basic idea is to describe a configuration problem (the problem domain) by a set of logical sentences.
The task of finding a configuration is then reduced to the problem of finding a model for the logical sentences ---
this is the same approach as the one taken by Klein et al.\ \cite{DFKI-TM-95-01}.
From Gottlob et al.\ we take the idea to express the conditional existence of components in configurations via implication and existential quantifiers.
However, we use counting quantifiers for this, and these are already present in the work by McGuinness et al. (albeit used for a different purpose).
The main idea of \LoCo is that via these counting quantifiers we can enforce that each model of the configuration problem contains finitely many components only.

\subsection{Formal Basics}

Formally, \LoCo is based on a fragment of classical logic with equality interpreted as identity. 
This fragment is then extended with existential counting quantifiers.

{\noindent \bf Components\/:} Components are modelled as $n$-ary predicates $\comp(\cid,\vec x)$, with \cid the component's identifier, and $\vec x$ a vector of component attributes.
Components are of various kinds; we will denote individual kinds by $C_1, C_2$.

{\noindent \bf Typed Variables\/:} 
It is convenient to say that the different arguments of components have different types.
We will introduce one type \sort{Id} for each identifier of a component kind and also for each attribute type.
We assume that there are only finitely many different types in the configuration domain that are all mutually disjoint.
In our notation we will use typed variables in formulas.

We now show how these typed variables can be accommodated in classical first order logic ---- 
this is very similar to the reduction of many-sorted logic to classical FO (cf.\ e.g.\ \cite{Enderton}).
We first introduce unary predicates for each type (e.g. $\text{ID}$ for type \sort{Id}) and add domain partitioning axioms\/:

\begin{flalign*}
&(\forall x) \bigvee_{T\in\mathcal{TYPES}} T(x),\\
&(\forall x) \bigwedge_{T_i,T_j\in\mathcal{TYPES},i\neq j} \neg (T_i(x) \land T_j(x)).
\end{flalign*}

Then for transforming a typed formula to an untyped one we replace 
e.g.\ each subformula $(\forall \cid) \phi(\cid)$ by $(\forall x) \text{ID}(x) \Rightarrow \phi(x)$ and likewise $(\exists \cid) \phi(\cid)$ by $(\exists x) \text{ID}(x) \land \phi(x)$ ---
this is the standard reduction from many-sorted to classical FO.
However, for the moment we are not going to introduce types for the terms (other than variables) of the language.
Later we are going to stipulate that there are {\em standard names} for the elements in the domain of each type, cf.\ Section~\ref{sec:cp}. 

{\noindent \bf Counting Quantifiers\/:}
For restricting the number of potential connections between components we use existential counting quantifiers $\cex$ 
with lower and upper bounds $l$ and $u$ such that $l\leq u$, $l \geq 0$ and $u > 0$. 
For example, we might have a formula $\cex x\phi(x)$ enforcing that the number of different $x$ (here $x$ denotes a sequence of variables) such that $\phi(x)$ is restricted to be within the range $[l,u]$. 
In classical logic without counting quantifiers this can be expressed as

\[
\bigvee_{l\leq n\leq u} \big[(\exists x_1, x_2, \ldots, x_n) [\phi(x_1) \land \phi(x_2) \land \ldots \land \phi(x_n)] \land [\bigwedge_{i\neq j}x_i\neq x_j] \land [(\forall x) \phi(x) \rightarrow \bigvee_i x = x_i\ ] \big].
\]

As usual quantifiers range over a single type only.
But occasionally, by an abuse of notation we will write e.g.\ $\cex x \phi(x) \lor \psi(x)$, where $\phi$ and $\psi$ expect different types.
This abbreviates a formula enforcing that the total number of objects such that $\phi$ or $\psi$ is between $l$ and $u$, where the disjunction is inclusive.
We denote exclusive disjunction between types in these subformulas by $\xcex x \phi(x) \lor \psi(x)$ ---
this abbreviates a formula enforcing that the total number of objects such that $\phi$ is between $l$ and $u$ and there are no $x$ such that $\psi$ (or the other way around).

{\noindent \bf Connections\/:} Configuration is about connecting components\/: For every set $\{C_1,C_2\}$ of potentially connected
components we introduce one of the binary predicate symbols $C_12C_2$ and $C_22C_1$ - it does not matter which. 
We allow connections from a component type to itself, i.e.,\ $C2C$.
A predicate $C_i2C_j$ is of type $\sort{Id}_i\times\sort{Id}_j$.
For every connection predicate $C_12C_2$ two formulas are included\/:\footnote{Throughout this paper free variables in formulas are to be read as existentially quantified from the outside.}

\begin{flalign}
\label{math:binary1}
&(\forall \cid_1,\vec x) \ C_1(\cid_1,\vec x) \Rightarrow \\ 
& \quad \quad \quad   (\exists^{u_{1}}_{l_{1}} \cid_2)\ C_12C_2(\cid_1,\cid_2) \land C_2(\cid_2,\vec y) \land \phi(\cid_1,\cid_2,\vec x, \vec y) \notag \\ \notag \\
\label{math:binary2}
&(\forall \cid_2,\vec x)\ C_2(\cid_2,\vec x) \Rightarrow \\ 
& \quad \quad \quad (\exists^{u_{2}}_{l_{2}} \cid_1)\ C_12C_2(\cid_1,\cid_2) \land C_1(\cid_1,\vec y) \land \psi(\cid_1,\cid_2,\vec x, \vec y) \notag
\end{flalign}

The first formula says how many components of kind $C_2$ can be connected to any given component of kind $C_1$,
with the subformula $\phi$ (with variables among $\cid_1,\cid_2,\vec x, \vec y$) expressing additional constraints, 
like e.g.\ an aggregate function $\sum n \leq \textsl{Capacity}$.
The second formula is for the other direction. 
If the connection is from a component kind to itself only one of the formulas is included.

The formulas $\phi$ and $\psi$ for expressing constraints on the connections consist of conjunctions and disjunctions of linear arithmetic expressions and (in-)equalities between terms.
We believe this to be sufficient for many practical examples; if necessary we will broaden the language, but we have to keep in mind the planned translation to executable formats.\footnote{For an explicit model of ports in \LoCo we can introduce attribute types for the ports and a binary predicate $\text{ConnectionPorts}$ that is then used in $\phi$. We do not do so here in order to simplify the presentation.}

Next to the rules for binary connections, there are also rules for supporting one-to-many connections (\ref{math:one-to-many}), 
i.e. connecting one component with a set of components. 
For every one-to-many connection the component on the left-hand side needs to have binary connections to all components in the set on the right-hand side. This is mandatory for the propagation of bounds and will be discussed later on. 
Note also that the single component is not allowed to be part of the set. 

\begin{flalign}
\label{math:one-to-many}
& (\forall \cid,\vec x) C(\cid,\vec x) \Rightarrow  (\cex \cid_i)\ [ \bigvee_i C2C_i(\cid,\cid_i) \land C_i(\cid_i,\vec y) ]
\end{flalign}

In this rule the quantifier \cex ranges over the $i>1$ different \sort{Id} types. 
It may also be replaced by the \xcex quantifier enforcing that each $C_1$ is connected to components of only one of the $C_i$ kinds. 
The cardinality upper bound is optional and in combination with the binary connections, a sufficient bound can be automatically computed. 

\subsection{Specifying Configuration Problems}
\label{sec:cp}

The specification \confprob of a configuration problem in our logic consists of two parts\/:

\begin{itemize}
\item domain knowledge in the form of the connection axioms, naming schemes, a component catalogue and an axiomatisation of arithmetic; and
\item instance knowledge in the form of component domain axioms.
\end{itemize}

Below we will speak of {\em input} and {\em generated} components. The intuition is that for the former we know exactly how many are used in a configuration and for the latter we don't.
We stipulate that a configuration problem always includes at least one component of the input variant. 

\subsubsection{Domain Knowledge}

{\noindent \bf Connection Axioms}
Connection axioms take the form introduced above.
Only in binary connection rules we allow the lower bound to be zero in the \cex quantifier, i.e.\ we can have $l=0$.
Without further conditions this would allow us to include infinitely many components into configurations\/:
Assume we have two components $C_1$ and $C_2$, where each $C_1$ is connected to exactly one $C_2$, and each $C_2$ is connected to at most one $C_1$.
It does not help if we know exactly how many $C_1$ there are (say $n$)\/: Still we can have infinitely many $C_2$ that are not connected to any of the $C_1$.

We address this problem as follows\/:
First, the component kinds have to be divided into the classes {\em input}, {\em generated} and {\em both}.
Then we stipulate that for every rule for binary connections from $C_1$ to $C_2$ with a lower bound of zero\/: 
$C_1$ is {\em input}, or there is some other binary or one-to-many connection from $C_1$ with lower bound greater than zero. 
Then we define a level mapping on the component kinds via the connection axioms\/:
Input components are on level zero. 
On level one are those generated components for which there is a (binary or one-to-many) connection axiom with lower bound greater than zero from the component to only input components.
Level two components are grounded in input or level one components, and so on.

Now any domain knowledge axiomatisation has to fulfil the following property\/:
No matter how the subdivision of component kinds into the classes {\em input}, {\em generated} and {\em both} is instantiated there has to exist a level mapping of the components such that
all components of the {\em generated} variety are assigned to some finite level. 
The existence of such a level mapping can be checked by first assigning {\em generated} to the components belonging to the class {\em both} and 
then doing a graph traversal starting from the input components.

\noindent{\bf Attribute Naming}
For all attribute types a naming-scheme is included.
For ordinary component attributes these take the form~(\ref{closed}) where $T$ is the unary type-predicate for the given type and \Vmc is a finite set of ground terms, the possible attribute values\/: 

\begin{flalign}
\label{closed}
& (\forall x)\ T(x) \equiv \bigvee_{V\in\Vmc}x = V
\end{flalign}

For component attributes of type \sort{Id} the naming-scheme has the form~(\ref{open}) where $\phi(x)$ is a FO formulation of the (infinitely many) possible names of elements in that type. For example, this could be a simple numbering axiom of form\/: $(\forall x) S(x) \Rightarrow (\exists n) x = \text{SName}(n)$. 

\begin{flalign}
\label{open}
& (\forall x)\ ID(x) \Rightarrow \phi(x),
\end{flalign}

By default, unique name axioms for all distinct terms are also included. 
Hence naming-scheme axioms of the form~(\ref{closed}) force the domain of the type to be equal to the set of all terms $t$ such that $\phi(t)$,
whereas the form~(\ref{open}) only forces the domain of the type to be a subset thereof. 

To sum up, for each component kind the \sort{Id} attribute is unbounded, but ordinary attributes can have only finitely many distinct values.
However, in each model of a configuration problem only finitely many components will exist. We introduce a new variable type \sort{Excess} without naming-scheme axiom\/:
The names of components not used in a configuration can be discarded by assigning them to this type.
Finally, for every component kind we introduce an axiom

\[
(\forall id_i,id_j,\vec x \vec y)[\ C(id_i,\vec x) \land C(id_j,\vec y) \land id_i = id_j\ ] \Rightarrow \vec x = \vec y
\]

expressing the fact that, in database terminology, the respective \sort{Id} is a {\em key}.

\noindent{\bf Component Catalogue}
For each component kind the so-called catalogue contains information on the instances that actually can be manufactured.
We express this as axioms (where each $\vec V_i$ is a tuple of ground attribute values)\/:

\[
(\forall id,\vec x) C(id,\vec x) \equiv \bigvee_i \vec x = \vec V_i
\]

\subsubsection{Instance Knowledge}
On the instance level the components assigned to the class {\em both} have to be divided into input and generated components.
For components $C$ of the input variant we make a closure assumption on the domain of the components identifiers\/:

\[
(\forall x) \text{ID}(x) \equiv \bigvee_{\text{ID}_i \in \mathcal{ID}} x = \text{ID}_i.
\]

where $\mathcal{ID}$ is a finite set of identifiers $\text{ID}_i$ and $\text{ID}$ is the respective type predicate.
This axiom is stronger than the naming-scheme for the component;
hence, in any model identifiers mentioned in the naming-scheme axiom but not in the domain closure axiom will belong to the type \sort{Excess}.

Both input and generated components that have to be used in the configuration can be explicitly listed as atoms (with possibly uninstantiated, existentially quantified arguments).
We also allow positive and negative ground connection predicates like, for example, $\neg C_12C_2(ID_1,ID_2)$.

\subsection{Finite Model Property}

Next we are going to show that in any model of a configuration domain specification for all components the domain of the \sort{Id} attribute is finite.

\begin{proposition}[Configurations contain finitely many components]
Let \confprob be a configuration domain specification and $\interpretation$ be an interpretation such that $\interpretation \models \confprob$. Then for all components 
the domain of the \sort{Id} attribute is finite in \interpretation.
\end{proposition}

\noindent{\bf Proof Sketch\/:}
Assume that for a component $C$ the domain of the \sort{Id} attribute is infinite and that $C$ is connected to some other component(s) $C_i$ via a binary or one-to-many connection
such that the domain of all the $C_i$ is finite in \interpretation. Then \interpretation is not a model for \confprob. The existence of a level mapping guarantees that each component
is grounded in components with finite domains.$\dashv$

{\noindent \bf Calculating upper and lower bounds\/:} 
In order to be able to transform a problem model into e.g.\ \SAT or OPL, we need to know the lower and upper bounds on the number of instances for each component of the ``generated'' variety. 
For computing these possible domain sizes of generated components, we extract Diophantine inequalities from the connection formulas.  
This builds up on the work by Falkner et al. about semantics of UML class diagrams and cardinalities applied to the configuration domain \cite{falkner:uml2}. 

Assume a binary connection defined by formulas (\ref{math:binary1}) and (\ref{math:binary2}), where $C_1$ is an input and $C_2$ is a generated component. We can calculate upper and lower bounds 
for component $C_{2}$ as follows\/:

\begin{flalign} 
\label{math:bounds1}
l_{1} * \left| C_1 \right| \leq n \leq u_{1} * \left| C_1 \right| \\
l_{2} * \left| C_2 \right| \leq n \leq u_{2} * \left| C_2 \right| \notag\\
\notag \\
\label{math:bounds2}
l_{1} * \left| C_1 \right| \leq u_{2}  * \left| C_2 \right| \\
l_{2} * \left| C_2 \right| \leq u_{1} * \left| C_1 \right| \notag
\end{flalign}

The number of possible links $n$ between the components is bounded as shown in (\ref{math:bounds1}). 
From this we can derive inequalities representing the relation between $C_1$ and $C_2$ (\ref{math:bounds2}). 
After some simple combinatorics we get lower bound $LB = \left\lceil \frac{ l_{1} * \left| C_{1} \right| }{ u_{2} } \right\rceil$ and 
upper bound $UB = \left\lfloor \frac{ u_{1} * \left| C_{1} \right| }{ l_{2} } \right\rfloor$, resulting in formula (\ref{math:bounds3}) for the bounds of $C_2$. It can be seen from the formula that to define a lower resp. an upper bound for $C_{2}$, we need the cardinality bounds $l_{2}$ resp. $u_{2}$ in the direction of $C_{1}$. 
The described computation also applies to connections between two generated components, provided that component $C_1$ has properly defined bounds. 
In this scenario we insert the lower bound on $C_1$ for computing $LB$ and the upper bound on $C_1$ for computing $UB$ of $C_2$. 

\begin{flalign}
\label{math:bounds3}
\left\lceil \frac{ l_{1} * \left\lfloor \left| C_{1} \right| \right\rfloor }{ u_{2} } \right\rceil \leq \left| C_{2} \right| \leq \left\lfloor \frac{ u_{1} * \left\lceil \left| C_{1} \right| \right\rceil }{ l_{2} } \right\rfloor 
\end{flalign}

\begin{flalign}
\label{math:bounds4}  
\left\lceil \frac{ \sum\limits_{i} l_{i} * \left\lfloor \left| C_{i} \right| \right\rfloor }{ u } \right\rceil \leq \left| C \right| \leq \left\lfloor \frac{ \sum\limits_{i} u_{i} * \left\lceil \left| C_{i} \right| \right\rceil }{ l } \right\rfloor 
\end{flalign}

In the case of one-to-many connections as shown in formula \ref{math:one-to-many} new bounds are calculated for the component on the left-hand side. 
For this computation we combine a one-to-many connection with all existing binary connections between the current component and the components on the many-side. 
In other words, we take the cardinalities from a one-to-many constraint in direction to the set and the cardinalities of the binary connections in direction to the current component and compute bounds analogously to a simple binary connection (see formula \ref{math:bounds4}). 

The above procedures refine the bounds on the domain of components in single connections. 
However, if the domain size of one component is updated, then the domain size of other components may have to be updated again.
We also have to take into account the one-to-many connections. 
The algorithm introduced below addresses both tasks until eventually for the domain sizes a fixpoint is obtained --- or a contradiction has been detected.

First, the algorithm puts all input components on a stack in any order (line~\ref{li:stack}). 
The algorithm then takes a component off the stack and iteratively determines all binary connections including the current component (line~\ref{li:binary-comps}). 
We perform a bound computation for each binary connection with a generated component and check if new bounds were computed for this generated component (line~\ref{li:new-bounds}). 
If this is the case, we update the bounds, taking the maximum of all computed lower bounds and the minimum of all computed upper bounds. 
After an update we check the bounds for consistency, i.e. we check that lower bound $\leq$ upper bound (line~\ref{li:consistency}) 
and put the connected component on the stack for further propagation of bounds (line~\ref{li:push1}). 
The algorithm terminates with an error whenever bounds become inconsistent. 

If the current component is of type generated, then we also check if there exist one-to-many connections to a set of other components and 
iterate over them (line~\ref{li:1-to-many-comps}). 
In the case of one-to-many connections new bounds are calculated for the current component and not for the connected components. 
If we obtain new bounds for the current component, we perform an update and a consistency check similar to what is done for the binary connections 
and put the component back on the stack again to propagate the new bounds via the binary connections. 

The algorithm then iteratively pops the next component off the stack and does the same computation step until the stack is empty. 
The algorithm is guaranteed to terminate and ensures the proper computation of maximal lower and minimal upper bounds on all generated components. 

\begin{codebox}
\Procname{$\proc{BOUND-PROPAGATION}$}
\li create an empty \id{stack}		\label{li:1}
\li put all $\id{inpComp} \in \const{input-components}$ on the $\id{stack}$ in any order \label{li:stack}
\li \While \id{stack} is not empty
\li \Do 
				$\id{currComp} \gets \func{POP}(\id{stack})$
\li			\For \textbf{all} $\left\langle \id{currComp}, \id{nbComp} \right\rangle \in \const{binary-connections}$ \label{li:binary-comps}
				\li \Do
								\If $\id{nbComp} \in \const{generated-components}$
\li									\Then
												$\proc{COMPUTE-BOUNDS}(\id{currComp}, \id{nbComp})$
								\li 		\If $\func{NEW-BOUNDS}(\id{nbComp})$ \label{li:new-bounds}
								\li     		\Then 
																$\proc{UPDATE-BOUNDS}(\id{nbComp})$
\li															\If $\func{LB}(\id{nbComp}) \leq \func{UB}(\id{nbComp})$ \label{li:consistency}
\li																	\Then
	 																			$\func{PUSH}(\id{nbComp}, \id{stack})$ \label{li:push1}
\li																	\Else
																				{\bf REJECT}
																		\End
												\End
										\End
						\End
\li			\If $\id{currComp} \in \const{generated-components}$
\li					\Then
								\For \textbf{all} $\left\langle \id{currComp}, \id{nbComps} \right\rangle \in \const{1-to-many-connections}$ \label{li:1-to-many-comps}
\li									\Do
												$\proc{COMPUTE-BOUNDS}(\id{currComp}, \id{nbComps})$
\li											\If $\func{NEW-BOUNDS}(\id{currComp})$
\li													\Then
																$\proc{UPDATE-BOUNDS}(\id{currComp})$
\li															\If $\func{LB}(\id{currComp}) \leq \func{UB}(\id{currComp})$
\li																	\Then
	 																			$\func{PUSH}(\id{currComp}, \id{stack})$
\li																	\Else
																				{\bf REJECT}
																		\End
												\End
										\End
						\End
		\End
\li {\bf ACCEPT}		
\end{codebox}

\section{Example: Modified Bin-Packing}

We want to explain our approach by means of a simple Bin-Packing example, where we distinguish between two component kinds of \textit{Things} \textit{A} and \textit{B} with all \textit{Things} having a certain size. 
The \textit{Bins} have an upper bound on how many \textit{Things} of each kind can be put into them. 
\textit{Things} are input components while the \textit{Bins} are generated components with the aim of their number being minimised. 
The problem can be described by the following formulas: 

\begin{flalign} 
\label{math:ex1}
& \forall(\cid_{\text{TA}}, size) \ C_{\text{TA}}(\cid_{\text{TA}}, size) \Rightarrow \\ 
& \quad \quad \quad (\exists^{1}_{1} \ \cid_{Bin}) \ C_{\text{TA}}2C_{Bin}(\cid_{\text{TA}},\cid_{Bin}) \land C_{Bin}(\cid_{Bin}) \notag \\ \notag \\
\label{math:ex2}
& \forall(\cid_{Bin}) \ C_{Bin}(\cid_{Bin}) \Rightarrow \\ 
& \quad \quad \quad (\exists^{5}_{0} \ \cid_{\text{TA}}) \ C_{\text{TA}}2C_{Bin}(\cid_{\text{TA}},\cid_{Bin}) \land C_{\text{TA}}(\cid_{\text{TA}}, size) \land \sum size \leq 5 \notag \\ \notag \\
\label{math:ex3}
& \forall(\cid_{\text{TB}}, size) \ C_{\text{TB}}(\cid_{\text{TB}}, size) \Rightarrow \\ 
& \quad \quad \quad ( \exists^{1}_{1} \ \cid_{Bin}) \ C_{\text{TB}}2C_{Bin}(\cid_{\text{TB}},\cid_{Bin}) \land C_{Bin}(\cid_{Bin}) \notag \\ \notag \\
\label{math:ex4}
& \forall(\cid_{Bin}) \ C_{Bin}(\cid_{Bin}) \Rightarrow \\ 
& \quad \quad \quad ( \exists^{2}_{0} \ \cid_{\text{TB}}) \ C_{\text{TB}}2C_{Bin}(\cid_{\text{TB}},\cid_{Bin}) \land C_{\text{TB}}(\cid_{\text{TB}}, size) \land \sum size \leq 2 \notag \\ \notag \\
\label{math:ex5}
& \forall(\cid_{Bin}) \ C_{Bin}(\cid_{Bin}) \Rightarrow \\
& \quad \quad \quad ( \exists^{}_{1} \ \cid_T) \ ( C_{\text{TA}}2C_{Bin}(\cid_{T},\cid_{Bin}) \land C_{\text{TA}}(\cid_{T},\vec y) ) \ \vee \notag \\
& \quad \quad \quad \quad \quad \quad \ \ \  ( C_{\text{TB}}2C_{Bin}(\cid_{T},\cid_{Bin}) \land C_{\text{TB}}(\cid_{T},\vec y) ) \notag
\end{flalign}
                                   
Formula (\ref{math:ex1}) states that every \textit{ThingA} has to be put into exactly one \textit{Bin}. 
The backwards-direction in formula (\ref{math:ex2}) determines that a \textit{Bin} has a total size bound of 5 for \textit{ThingA}. 
Up to 5 of those things can be put into a \textit{Bin} in case all those \textit{Things} have minimum size 1 (hence the cardinality upper bound is 5). 
Formulas (\ref{math:ex3}) and (\ref{math:ex4}) analogously define the binary connection for \textit{ThingB}. 

Assume having an instance with 20 \textit{Things} of each kind, connection \textit{ThingA-Bin} gives a lower bound of 4 and connection \textit{ThingB-Bin} gives a lower bound of 10 for component \textit{Bin} using the bound computations defined in (\ref{math:bounds3}). We take the maximum of all available values, hence the lower bound for \textit{Bin} is 10. 
Notice that in (\ref{math:ex2}) and (\ref{math:ex4}) the cardinality lower bounds of the connections are defined as zero to express the situation that a \textit{Bin} could contain only one kind of \textit{Thing} without the other. 
This results in the fact that we can't compute an upper bound for \textit{Bin} using the binary connections defined so far and this would violate the finite model requirement. 
In order to express that for a \textit{Bin} to exist it needs to have at least one \textit{Thing} in it, we define a one-to-many connection between \textit{Bin} and the set of \textit{Things} (\ref{math:ex5}). It is sufficient to only define a lower bound for this connection and in conjunction with the binary connections we can now compute an upper bound of 40 for a \textit{Bin}, which would occur in a situation where every 
\textit{Thing} would be put in a separate \textit{Bin}. 

\section{Conclusion and Future Work}

We presented the core of \LoCo, a high-level language for modelling configuration problems, including the conditional generation of components. 
The key feature of the formalism is that the number of components used in configurations is bounded implicitly by the possible number of connections between components.
As a next step we plan to extend \LoCo so that it is possible to\/:

\begin{itemize}
\item express that the presence of one connection in a configuration depends on the presence of some other connection;
\item specify arbitrary combinations of components in the rules for one-to-many connections; and
\item incorporate a component taxonomy, where components can be subkinds of other components.
\end{itemize}

Once this is completed we plan to translate \LoCo to an executable format such as \SAT, OPL or answer set solving.
We also intend to carefully analyse the complexity of e.g.\ model construction or the bounds propagation algorithm in \LoCo.

{\noindent \bf Acknowledgement} We thank Markus Stumptner and Heribert Vollmer for many helpful discussions.

\bibliographystyle{eptcs}
\bibliography{references}

\begin{thebibliography}{10}
\providecommand{\bibitemdeclare}[2]{}
\providecommand{\urlprefix}{Available at }
\providecommand{\url}[1]{\texttt{#1}}
\providecommand{\href}[2]{\texttt{#2}}
\providecommand{\urlalt}[2]{\href{#1}{#2}}
\providecommand{\doi}[1]{doi:\urlalt{http://dx.doi.org/#1}{#1}}
\providecommand{\bibinfo}[2]{#2}

\bibitemdeclare{techreport}{DFKI-TM-95-01}
\bibitem{DFKI-TM-95-01}
\bibinfo{author}{Martin Buchheit}, \bibinfo{author}{R{\"u}diger Klein} \&
  \bibinfo{author}{Werner Nutt} (\bibinfo{year}{1995}):
  \emph{\bibinfo{title}{Constructive Problem Solving: {A} Model Construction
  Approach towards Configuration}}.
\newblock \bibinfo{type}{Technical Report} \bibinfo{number}{TM-95-01},
  \bibinfo{institution}{DFKI}.
\newblock
  \urlprefix\url{ftp://ftp.dfki.uni-kl.de/pub/Publications/TechnicalMemos/1995/TM-95-01.ps.gz}.

\bibitemdeclare{book}{Enderton}
\bibitem{Enderton}
\bibinfo{author}{H.~B. Enderton} (\bibinfo{year}{1972}):
  \emph{\bibinfo{title}{A Mathematical Introduction to Logic}}.
\newblock \bibinfo{publisher}{Academic Press}.

\bibitemdeclare{article}{falkner:uml2}
\bibitem{falkner:uml2}
\bibinfo{author}{Andreas Falkner}, \bibinfo{author}{Ingo Feinerer},
  \bibinfo{author}{Gernot Salzer} \& \bibinfo{author}{Gottfried Schenner}
  (\bibinfo{year}{2010}): \emph{\bibinfo{title}{Computing Product
  Configurations via {UML} and Integer Linear Programming}}.
\newblock {\sl \bibinfo{journal}{Journal of Mass Customisation}}
  \bibinfo{volume}{3}(\bibinfo{number}{4}), pp. \bibinfo{pages}{351--367},
  \doi{10.1504/IJMASSC.2010.037650}.

\bibitemdeclare{inproceedings}{FriedrichStumptnerConfWS99}
\bibitem{FriedrichStumptnerConfWS99}
\bibinfo{author}{Gerhard Friedrich} \& \bibinfo{author}{Markus Stumptner}
  (\bibinfo{year}{1999}): \emph{\bibinfo{title}{Consistency-Based
  Configuration}}.
\newblock In: {\sl \bibinfo{booktitle}{Configuration Workshop at AAAI'99}}.

\bibitemdeclare{inproceedings}{gottlob:conditional}
\bibitem{gottlob:conditional}
\bibinfo{author}{Georg Gottlob}, \bibinfo{author}{Gianluigi Greco} \&
  \bibinfo{author}{Toni Mancini} (\bibinfo{year}{2007}):
  \emph{\bibinfo{title}{Conditional Constraint Satisfaction: Logical
  Foundations and Complexity}}.
\newblock In: {\sl \bibinfo{booktitle}{IJCAI'07}}.

\bibitemdeclare{incollection}{junker:configuration}
\bibitem{junker:configuration}
\bibinfo{author}{Ulrich Junker} (\bibinfo{year}{2006}):
  \emph{\bibinfo{title}{Configuration}}.
\newblock In \bibinfo{editor}{F.~Rossi}, \bibinfo{editor}{P.~van Beek} \&
  \bibinfo{editor}{T.~Walsh}, editors: {\sl \bibinfo{booktitle}{Handbook of
  Constraint Programming}}, \bibinfo{publisher}{Elsevier}, pp.
  \bibinfo{pages}{837 -- 874}.

\bibitemdeclare{inproceedings}{kolaitis:cqc_csp}
\bibitem{kolaitis:cqc_csp}
\bibinfo{author}{Phokion~G. Kolaitis} \& \bibinfo{author}{Moshe~Y. Vardi}
  (\bibinfo{year}{1998}): \emph{\bibinfo{title}{Conjunctive-Query Containment
  and Constraint Satisfaction}}.
\newblock In: {\sl \bibinfo{booktitle}{PODS'98}}, \doi{10.1145/275487.275511}.

\bibitemdeclare{article}{McguinnessW98}
\bibitem{McguinnessW98}
\bibinfo{author}{Deborah~L. McGuinness} \& \bibinfo{author}{Jon~R. Wright}
  (\bibinfo{year}{1998}): \emph{\bibinfo{title}{Conceptual modelling for
  configuration: {A} description logic-based approach}}.
\newblock {\sl \bibinfo{journal}{AI EDAM}}
  \bibinfo{volume}{12}(\bibinfo{number}{4}), pp. \bibinfo{pages}{333--344},
  \doi{10.1017/S089006049812406X}.

\bibitemdeclare{inproceedings}{mittal:dcsp}
\bibitem{mittal:dcsp}
\bibinfo{author}{S.~Mittal} \& \bibinfo{author}{B.~Falkenhainer}
  (\bibinfo{year}{1990}): \emph{\bibinfo{title}{Dynamic constraint satisfaction
  problems}}.
\newblock In: {\sl \bibinfo{booktitle}{AAAI'90}}.

\bibitemdeclare{inproceedings}{mittal:generic_model}
\bibitem{mittal:generic_model}
\bibinfo{author}{Sanjay Mittal} \& \bibinfo{author}{Felix Frayman}
  (\bibinfo{year}{1989}): \emph{\bibinfo{title}{Towards a Generic Model of
  Configuration Tasks}}.
\newblock In: {\sl \bibinfo{booktitle}{IJCAI'89}}.

\bibitemdeclare{inproceedings}{sabin:composite}
\bibitem{sabin:composite}
\bibinfo{author}{Daniel Sabin} \& \bibinfo{author}{Eugene~C. Freuder}
  (\bibinfo{year}{1996}): \emph{\bibinfo{title}{Configuration as Composite
  Constraint Satisfaction}}.
\newblock In: {\sl \bibinfo{booktitle}{AIMRP'96}}.

\bibitemdeclare{article}{sabin:survey}
\bibitem{sabin:survey}
\bibinfo{author}{Daniel Sabin} \& \bibinfo{author}{Rainer Weigel}
  (\bibinfo{year}{1998}): \emph{\bibinfo{title}{Product Configuration
  Frameworks - A Survey}}.
\newblock {\sl \bibinfo{journal}{IEEE Intelligent Systems}}
  \bibinfo{volume}{13}(\bibinfo{number}{4}), pp. \bibinfo{pages}{42--49},
  \doi{10.1109/5254.708432}.

\bibitemdeclare{article}{stumptner:generative}
\bibitem{stumptner:generative}
\bibinfo{author}{M.~Stumptner}, \bibinfo{author}{A.~Haselb\"{o}ck} \&
  \bibinfo{author}{G.~Friedrich} (\bibinfo{year}{1998}):
  \emph{\bibinfo{title}{Generative constraint-based configuration of large
  technical systems}}.
\newblock {\sl \bibinfo{journal}{AI EDAM}}
  \bibinfo{volume}{12}(\bibinfo{number}{4}), pp. \bibinfo{pages}{307--320},
  \doi{10.1017/S0890060498124046}.

\bibitemdeclare{inproceedings}{thorstensen:capturing}
\bibitem{thorstensen:capturing}
\bibinfo{author}{Evgenij Thorstensen} (\bibinfo{year}{2010}):
  \emph{\bibinfo{title}{Capturing configuration}}.
\newblock In: {\sl \bibinfo{booktitle}{Doctoral Program at CP'10}}.

\end{thebibliography}

\end{document}